\documentclass{pasj00}
\draft

\begin{document}
\SetRunningHead{S. Hasegawa et al.}{Asteroid catalog from slow-scan observations}

\title{The Asteroid Catalog Using AKARI IRC Slow-Scan Observations}

 \author{%
   Sunao \textsc{Hasegawa},\altaffilmark{1}
   Thomas G. \textsc{M\"uller},\altaffilmark{2}
   Daisuke \textsc{Kuroda},\altaffilmark{3}
   Satoshi \textsc{Takita},\altaffilmark{1}
   and
   Fumihiko \textsc{Usui}\altaffilmark{1}}
 \altaffiltext{1}{Institute of Space and Astronautical Science, Japan Aerospace Exploration Agency, \\ 3-1-1 Yoshinodai, Sagamihara, Kanagawa 229-8510, Japan}
 \email{hasehase@isas.jaxa.jp}
 \altaffiltext{2}{Max-Planck-Institut f\"ur Extraterrestrische Physik, Giessenbachstra\ss e, 85748 Garching, Germany}
 \altaffiltext{3}{Okayama Astrophysical Observatory, National Astronomical Observatory, \\ 3037-5 Honjo, Kamogata-cho, Asakuchi, Okayama 719-0232, Japan}

\KeyWords{catalogs  ---  infrared: solar system  ---  minor planets, asteroids  ---  space vehicles: instruments} 

\maketitle

\begin{abstract}
We present an asteroidal catalog from the mid-infrared wavelength region using the slow-scan observation mode obtained by the Infrared Camera (IRC) on-board the Japanese infrared satellite {AKARI}.
An archive of IRC slow-scan observations comprising about 1000 images was used to search for serendipitous encounters of known asteroids.
We have determined the geometric albedos and diameters for 88 main-belt asteroids, including two asteroids in the Hilda region, and compared these, where possible, with previously published values.
Approximately one-third of the acquired data reflects new asteroidal information.
Some bodies classified as C or D-type with high albedo were also identified in the catalog.
\end{abstract}

\section{Introduction}
Physical studies of asteroids provide important information on the present conditions of the solar system.
In turn, this can be used to understand the early history and evolution of the solar system.
Key data in the physical study of asteroids are reliable geometric albedo and diameter measurements.
Geometric albedos place approximate constraints on the surface composition of asteroids \citep{Burbine2008}.
The distribution of the geometric albedos of main-belt asteroids provides insight into the chemical and mineralogical evolution of the solar system in the vicinity of the main-belt (Usui et al. ApJ submitted).
Combined with estimates of mass, asteroidal diameter can yield bulk density, which is diagnostic of the internal structure of asteroids \citep{Carry2012}.
The size distribution of asteroids can also reveal their original mass and the collisional history of the main-belt asteroids \citep{Bottke2005}.

The radiometric method for determining the geometric albedo and the diameter of asteroids has been used since the early 1970s \citep{Allen1970}, and is a precise method that enables the acquisition of a large amount of data.
For example, estimation of the diameter of asteroid (25143) Itokawa using thermal observations is in excellent agreement with subsequent in situ measurements \citep{Mueller2007}.
Previous studies using the dedicated observations have presented a large amount of data on the geometric albedo and diameter of asteroids, including 84 asteroids by \citet{Hansen1976}, 84 asteroids by \citet{Morrison1977}, 352 by \citet{Gradie1988},  44 by \citet{Fernandez2009}, and 101 by \citet{Trilling2010}.

The survey that covered the whole sky produced great benefits for study of the asteroids. 
The {Infrared Astronomical Satellite} ({IRAS}; \cite{Neugebauer1984}) was launched on January 26, 1983, and mapped \textgreater{}96\% of the sky in four wavelength bands during its 10-month mission life.
An unbiased catalog of geometric albedo and diameter data for asteroids was established from this all-sky survey  \citep{Matson1986}, which ultimately resulted in the geometric albedo and diameter of 2,470 asteroids being presented in the Supplemental {IRAS} Minor Planet Survey (SIMPS; \cite{Tedesco2002b}).
Approximately two decades after the {IRAS} all-sky survey, two infrared astronomical satellites, {AKARI} (previously known as ASTRO-F; \cite{Murakami2007}) and the {Wide-field Infrared Survey Explorer} ({WISE}; \cite{Wright2010}) performed all-sky surveys in the infrared wavelength region.
{AKARI} was launched on February 21, 2006 and its survey to cover the whole sky in six bands took place over an 18-month mission life.
{WISE} conducted mapping of the whole sky in four bands following its launch on December 14, 2009, which continued until refrigerant expired after eight months.
{AKARI} and {WISE} acquired geometric albedo and diameter data for 5,120 \citep{Usui2011} and more than 130,000 asteroids (\cite{Grav2011}; \cite{Grav2012}; \cite{Mainzer2011}; \cite{Masiero2011}), respectively.

The radiometric method can obtain geometric albedo and diameter information using not only dedicated observations and the all-sky survey data, but also serendipitous data.
The mission purpose of the {Midcourse Space Experiment} satellite ({MSX}; \cite{Mill1994}), launched on April 24, 1996, was the acquisition of a range of data including an engineering test, an investigation of the composition of Earth's atmosphere, and astronomical research.
The {MSX} infrared astronomy observations were performed to fill a gap in the {IRAS} all-sky survey observations \citep{Price2001}.
Using these data, the geometric albedo and diameter of 168 different asteroids were serendipitously obtained \citep{Tedesco2002a}.
The Infrared Space Observatory ({ISO}; \cite{Kessler1996}) was launched on November 17, 1995 and 
conducted \textgreater{}30,000 individual observations.
{ISO} also carried out parallel and serendipitous observations of the sky whilst the satellite was slewing from one target to the next.
The ISOPHOT Serendipity Survey (ISOSS; \cite{Bogun1996}) obtained 170 \micron~ sky brightness data, resulting in 31 asteroids being identified \citep{Mueller2002}.
The {Spitzer Space Telescope} ({SST}; \cite{Werner2004}), launched on August 25, 2003, was an observatory type mission like that of {ISO} and dedicated a significant part of its time to the Legacy programs.
Using archive data of the Legacy programs: the First Look Survey-Ecliptic Plane Component (FLS-EPC), the Great Observatories Origins Deep Survey (GOODS), a 24 and 70 \micron~ Survey of the Inner Galactic Disk with MIPS (MIPSGAL), the Spitzer Deep Survey of the HST COSMOS 2-Degree ACS Field (SCOSMOS), the Spitzer Wide-Area Infrared Extragalactic Survey (SWIRE), and Taurus, the serendipitous detection of many small asteroids were conducted (\cite{Lonsdale2003}, \cite{Meadows2004}, \cite{Sanders2007}, \cite{Ryan2009}, \cite{Bhattacharya2010}, Ryan et al. AJ submitted).
{AKARI} was designed as an all-sky survey mission in the infrared region, similar to the {IRAS} and {WISE} surveys. 
However, {AKARI} had the capability to make pointed observations in addition to its all-sky survey.
Approximately 5,000 pointed observations were performed in the cryogenic phase of the {AKARI} mission before the liquid helium on-board the satellite was exhausted.
{AKARI} is equipped with a 68.5 cm Ritchey-Chr\'{e}tien type telescope and two scientific instruments: the Far-Infrared Surveyor (FIS; \cite{Kawada2007}) and the Infrared Camera (IRC; \cite{Onaka2007}).
The IRC on-board {AKARI} was designed to carried out deep photometric imaging and spectroscopy in the pointed observation mode.
The IRC can also operate in a scanning operation readout mode that extends the all-sky survey wavelength coverage into the mid-infrared wavelength range \citep{Ishihara2010}.
The combination of the scanning operation readout mode and pointed observations allows IRC to make observations of large areas to moderate depth, which is referred to as IRC slow-scan observations \citep{Takita2012}.

Here we present a catalog of the geometric albedo and diameter data for asteroids serendipitously detected by IRC slow-scan observations.
We first describe the data processing and catalog creation in Section 2, which is followed by a discussion of the results in Section 3, and a summary of the main findings of this study in Section 4.

\section{Data Processing and Catalog Creation}
\subsection{IRC Slow-Scan Observations}
The IRC comprises three cameras (NIR, MIR-S, and MIR-L) that cover wavelength ranges of 2--5, 5--13, and 12--26 \micron, respectively.
The NIR and two MIR cameras have infrared sensor arrays of \timeform{512} $\times$ \timeform{412} and \timeform{256} $\times$ \timeform{256} pixels, and pixel sizes of ca. 1.5\timeform{"}/pixel and 2.4\timeform{"}/pixel, respectively.
The three cameras have a field of view of ca. \timeform{10'} $\times$ \timeform{10'}, and the NIR and MIR-S cameras share the same field of view through a beam splitter, whereas the MIR-L camera observes the sky at ca. \timeform{20'} away from the other cameras in a direction perpendicular to the {AKARI} scan direction \citep{Onaka2007}.

The two MIR cameras are able to operate in a scanning operation readout mode \citep{Ishihara2006}, which was developed for the IRC all-sky survey.
The NIR camera is not used during the scanning operation readout mode, as the alignment of the NIR array is not suited for all-sky survey observations.
The scanning operation readout mode is achieved by operating the MIR cameras in two lines of two-dimensional detector arrays in both continuous and non-destructive readout modes.
The sampling rate was set at one sampling per 44 ms of observation time.
The MIR-S and MIR-L cameras were operated with filters at \textit{S9W} (reference wavelength: 9.0 \micron) and \textit{L18W}(reference wavelength: 18.0 \micron) bands, which have effective bandwidths of 4.1 and 10.0 \micron~ in the scanning operation readout mode, respectively.

Although the scanning operation readout mode was developed for the IRC all-sky survey, this mode can also be used for pointed slow-scan observations \citep{Takita2012}.
The telescope scanned the sky with much slower speed (\timeform{8}, \timeform{15}, or \timeform{30"} $\mathrm{s^{-1}}$) than during the all-sky survey (\timeform{216"} $\mathrm{s^{-1}}$).
During slow-scan observations, the telescope scanned along one or two round-trip paths centered about the target position to acquire a redundant dataset.
Due to the telescope scan speed and number of round-trip paths, the field of view of IRC slow-scan observations in the scan direction changes from \timeform{21.5'} to \timeform{2.9D}, but perpendicular to the scan direction the field of view varies little (\timeform{9'}--\timeform{14'}).

IRC slow-scan observations are given in the Astronomical Observation Templates (AOTs) of IRC11 and IRC51.
The differences in the AOTs of IRC11 and IRC51 are only the method that enabled examination of the readout signals from the detectors.
Pixels were reset at every 51 and 306 samplings to discharge the photo-current for the AOTs of IRC51 and IRC11, respectively.
For the IRC11, the detector arrays were operated in the same manner as in the all-sky survey mode, involving binning of 4 pixels in the cross-scan direction.
Given that data acquisition of the second line in the array was started at half of the integration time of the first line, the virtual pixel size of IRC11 is 2 pixels \citep{Ishihara2010}.
Conversely, the AOT of IRC51 provides full spatial resolution in the cross-scan direction.
The full spatial resolution of the IRC with pixel size of \timeform{2.5"} in the cross-scan direction was abandoned for the AOT of IRC51, whereas the AOT of IRC11 had a pixel size of \timeform{5"} after pixel binning.

After the exhaustion of liquid helium on August 26, 2008, the MIR-S and MIR-L cameras could not continue to be operated.
IRC slow-scan observations from January 2007 were mainly executed with the AOT of IRC51; consequently, most of the IRC slow-scan observations were made with the AOT of IRC51.
Given that the FIS and IRC can be operated simultaneously, 580 IRC slow-scan data (ca. 50 data from the AOT of IRC11 and ca. 530 data from the AOT of IRC51) also include parallel observations from the FIS-dedicated observations.
The two cameras with filters at \textit{S9W} and \textit{L18W} acquired 1,040 images in IRC slow-scan observations.
All of the IRC slow-scan observations were not actually primarily acquired for the purpose of asteroid research, but rather for other fields of infrared astronomy research.
Therefore, only ca. \textless{}10\% of all the IRC slow-scan observations were in the direction of the ecliptic plane.

\subsection{Data Reduction Processes}
The data packages of the IRC slow-scan observations are stored in dedicated FITS file format, referred to as Time-Series Data (TSD), which were originally developed for the FIS \citep{Takita2012}.
The TSD is a binary FITS table and consists of signal output, house-keeping details, and attitude information data for the satellite.
Two FITS tables were created from one pointed observation from the MIR-S and MIR-L cameras of the IRC.

The IRC slow-scan observation data were reduced by AKARI data reduction tools (ARIS; \cite{Takita2012}), which are written in Interactive Data Language (IDL). 
Most of the ARIS processing can handle the TSD file formats.
The following basic calibrations were applied to the TSD format data in order to correct for the anomalous behavior of the detector output after the reset: non-linearity between incoming photons and output signals, data differentiation, dark subtraction, flat fielding, and masking of bad pixels.
Two-dimensional image data were constructed after execution of these basic corrections and calibrations.

IRC slow-scan observational images of the \textit{S9W} and \textit{L18W} in the AOT of IRC11 and \textit{S9W} and \textit{L18W} in the AOT of IRC51 after processing by ARIS have \timeform{4.6"}/pixel, \timeform{5.0"}/pixel, \timeform{2.3"}/pixel, and \timeform{2.5"}/pixel, respectively.
The full width at half maximum sizes of the point spread functions are ca. \timeform{8"} and \timeform{10"} for the AOT of IRC11 and ca. \timeform{6"} and \timeform{7"} for the AOT of IRC51 for the \textit{S9W} and the \textit{L18W} bands, respectively.

\subsection{Flux Calibration for Point Sources}
The conversion factors for the AOT of IRC51 were derived from relationships between the intensities in ADU from the images and the calculated in-band flux densities in units of Jy for standard stars given in \citet{Takita2012}.
Since IRC11 is different from IRC51 in the reading method from detectors (Section 2.1), the conversion coefficient of IRC11 should be examined separately.
However, \citet{Takita2012} did not derive conversion factors for the AOT of IRC11.
Using the same technique to obtain the conversion factors for the AOT of IRC51, the conversion factors for IRC11 were determined as follows.

Infrared observations of two standard stars \citep{Cohen1999} were used for the absolute flux calibrations of IRC slow-scan observations in the AOT of IRC11 (Table \ref{tab:std}).
Aperture photometry for the standard stars imaged with MIR-S and MIR-L in the AOT of IRC11 was performed using the APPHOT task of IRAF with a circular aperture radius of 5.2 and 5.3 pixels, respectively.
This enabled the sky values for an annulus outside the aperture with a width of 3.5 pixels for the \textit{S9W} band and 3.6 pixels for the \textit{L18W} band to be determined.

\begin{longtable}{lllcc}
  \caption{Observed standard stars for in the AOT IRC11}\label{tab:std}
  \hline              
  name & RA (J2000) & Dec (J2000) & \textit{S9W} in-band flux [Jy] & \textit{L18W} in-band flux [Jy] \\ 
\endfirsthead
  \hline
  name & \textit{S9W} & \textit{L18W} \\
\endhead
  \hline
\endfoot
  \hline
\endlastfoot
  \hline
  KF09T1        & 17:59:23.04 & +66:02:56.1 & 3.458$\times$$\mathrm{10^{-2}}$ & -- \\
  NPM1p67\_0536 & 17:58:54.66 & +67:47:36.8 & -- & 3.755$\times$$\mathrm{10^{-2}}$ \\
  HD42525       & 06:06:09.37 & $\mathrm{-}$66:02:22.7 & 2.799$\times$$\mathrm{10^{-1}}$   & 5.876$\times$$\mathrm{10^{-2}}$   \\
\end{longtable}

\begin{figure}
  \begin{center}
    \FigureFile(80mm,80mm){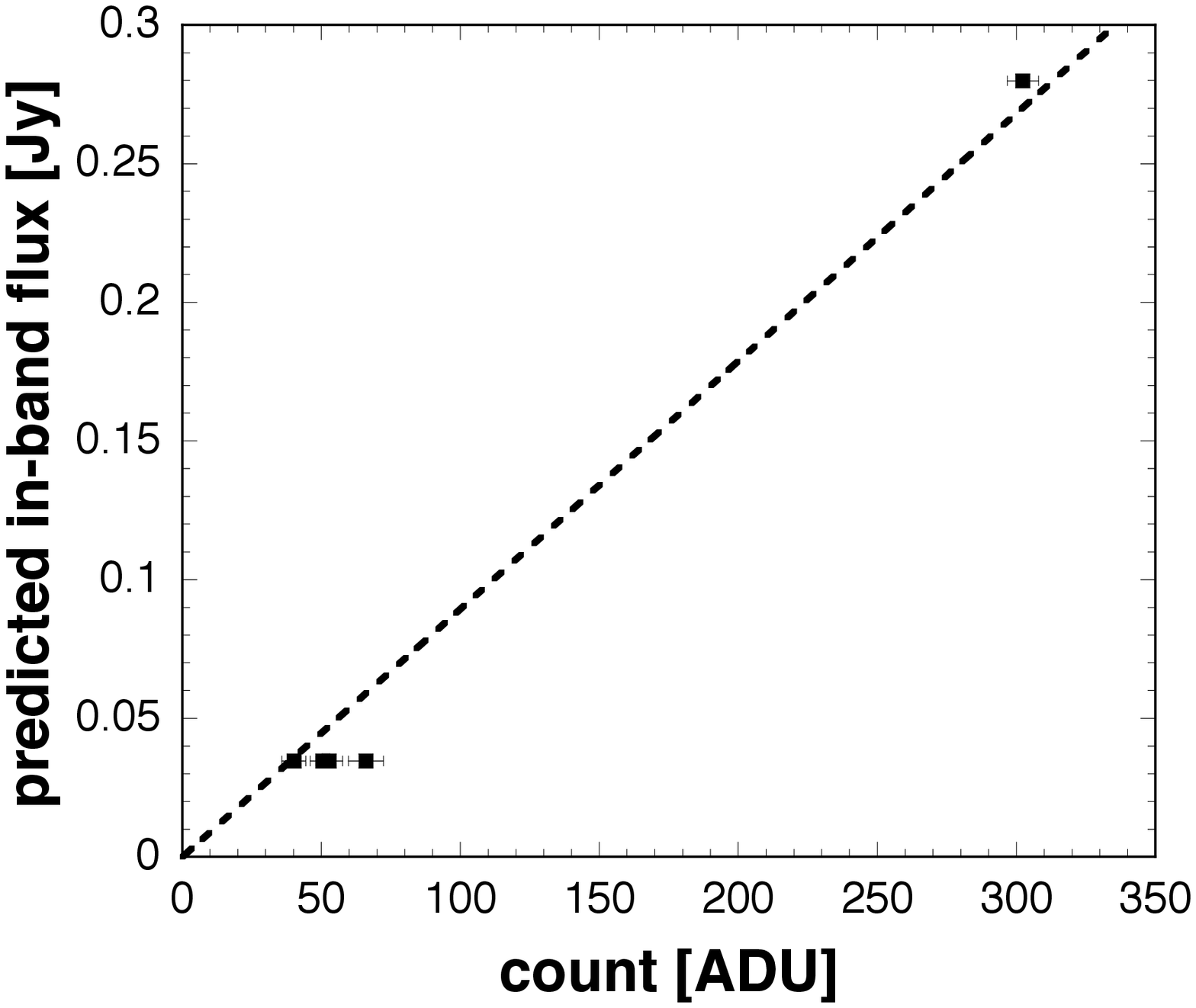}
    \FigureFile(80mm,80mm){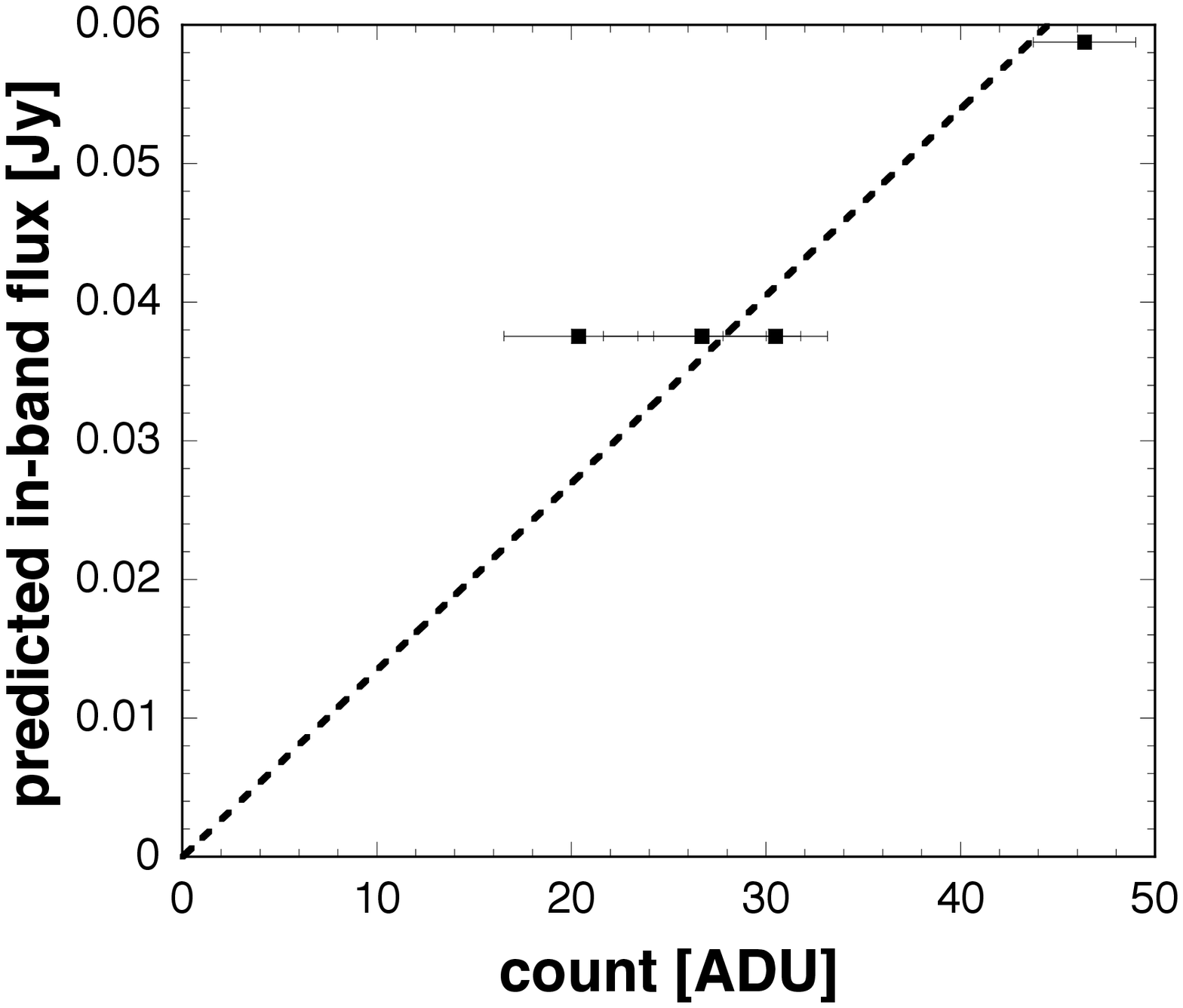}
  \end{center}
  \caption{Relationship between estimated flux density and observed counts of standard stars for the \textit{S9W} (left) and \textit{L18W} (right) bands. 
The dashed lines in each plot represent a least-squares linear regression of the data.
}
\label{fig:conversion factors}
\end{figure}

The in-band flux densities of the standard stars were taken from \citet{Tanabe2008}.
The intensities in ADU from the images in the AOT11 are shown in Figure \ref{fig:conversion factors} as a function of the predicted in-band flux of the standard stars. 
A method of the fitting is a method same as \citet{Takita2012}.
The slopes of the fitted lines in Figure \ref{fig:conversion factors} provide the conversion factors for \textit{S9W} and \textit{L18W}.
The conversion factors in units of Jy $\mathrm{ADU^{-1}}$ for the AOTs of IRC11 and IRC51 (Table 4 in \cite{Takita2012}) are listed in Table \ref{tab:conversionfactor}.

\begin{longtable}{lcc}
  \caption{Conversion factors for the IRC slow-scan observations [Jy $\mathrm{ADU^{-1}}$]}\label{tab:conversionfactor}
  \hline              
  Filter band \& AOT  & Conversion factors & Error \\ 
\endfirsthead
  \hline
  Band & Conversion factors & Error \\
\endhead
  \hline
\endfoot
  \hline
\endlastfoot
  \hline
  \textit{S9W} in the AOT11  & 8.942$\times$$\mathrm{10^{-4}}$   & 4.845$\times$$\mathrm{10^{-5}}$ \\
  \textit{S9W} in the AOT51  & 2.518$\times$$\mathrm{10^{-4}}$   & 3.228$\times$$\mathrm{10^{-6}}$ \\
  \textit{L18W} in the AOT11 & 1.328$\times$$\mathrm{10^{-3}}$   & 8.211$\times$$\mathrm{10^{-5}}$ \\
  \textit{L18W} in the AOT51 & 5.291$\times$$\mathrm{10^{-4}}$   & 1.365$\times$$\mathrm{10^{-6}}$ \\
\end{longtable}

\subsection{Asteroidal Search}
It is difficult to search for unknown asteroids given the lack of continuity of images obtained from the IRC slow-scan observations.
Therefore, a search for known asteroids was performed, involving application of N-body simulations including gravitational perturbations for the Moon, eight planets, (1) Ceres, (2) Pallas, (4) Vesta, and (134340) Pluto to the orbital calculations of asteroids with known orbital elements.
The position of the asteroids was calculated using the Runge--Kutta--Nystr\"om 12(10) method \citep{Dormand1987}.
The orbital elements for the asteroids used as inputs in the calculations were taken from the Asteroid Orbital Elements Database \citep{Bowell1994} of Lowell Observatory at the epoch of March 14, 2012.
This database contains 544,599 registered asteroids (326,254 numbered and 218,345 unnumbered). 
The orbital elements of the Sun, planets (including (134340) Pluto), and the Moon were taken from the DE405 JPL Planetary and Lunar Ephemerides in the J2000.0 equatorial coordinates from the NASA Jet Propulsion Laboratory.
The parallax between the geocenter and {AKARI} satellite is not negligible, because AKARI circles in a Sun-synchronous polar orbit at an altitude of 700 km.
The observing position of {AKARI} was adopted in this study, obtained by interpolation of data from the {AKARI} observational scheduling tool, which provides an observing position with sufficient accuracy for the purpose of this study.
Taking into account corrections for the light-time, gravitational deflection of light, stellar aberration, and precession and nutation of Earth's rotational axis, calculation of the astronomical coordinates in the search for known asteroids was carried out.
In addition, the heliocentric and "AKARI-centric" distance and the phase angle, which are necessary for the thermal models, were calculated.
These calculations are similar to those used for the Asteroid catalog using AKARI (AcuA; \cite{Usui2011}), which was based on the all-sky survey of AKARI.

Using the acquired IRC slow-scan observational images, information regarding the position, size of the field of view, and observed time can be checked.
Based on position and observed time data, known asteroids that should be present in a field of view of a particular image were searched for.
The calculated position of a known asteroid is then compared with the position of a point source in the image.
If a calculated position includes a point source within \timeform{10"}, the point source is considered to be a candidate for that the asteroid.
With regards to the thermal flux of the asteroid, a range of brightness is physically restricted in relation to the asteroid albedo, which is typically in the range of 0.01--1.00.
If the observed flux of the candidate asteroid matched expected flux range of the known asteroid, then the point source was identified as the asteroid in question.

\subsection{Asteroidal Flux Measurements}
The AKARI telescope is not able to track moving objects, such as comets and asteroids.
Thus, for moving objects, a centroid determination in combination with a standard shift-and-add technique was performed before stacking processing, in order to avoid compromising the photometric accuracy.
Aperture photometry for asteroids in the IRC slow-scan observation mode was carried out with the APPHOT task of IRAF using the same aperture radius and sky width as those used in the standard star flux calibrations (Section 2.3 and \cite{Takita2012}).

Color differences between the calibration stars (A-type main sequence and K-type giant stars) and asteroids need to be considered, given the wide bandwidth of \textit{S9W} and \textit{L18W}.
In order to obtain accurate color-corrected monochromatic flux data, color corrections were applied that were constrained by the shape of spectral energy distributions.
As the temperature on the surface of an asteroid changes with the heliocentric distance of the asteroid, the color correction factor is expressed as a polynomial function of the heliocentric distance of the object \citep{Usui2011}.
The color-corrected monochromatic flux $F_{\rm cc}$ in units of Jy, is given by:

\begin{eqnarray}
F_{\rm cc} &=& \frac{F_{\rm in\mathchar`-band}}{E_{\rm ccf}} ,
\label{eq:colour correction factors}
\end{eqnarray}
where $F_{\rm in\mathchar`-band}$ is the in-band flux in units of Jy. 
The color correction factor $E_{\rm ccf}$ in each filter band is derived from:

\begin{equation} 
E_{\rm ccf} = \left\{ \begin{array}{lcl}
0.984 - 0.0068 R_{\rm h} + 0.031 R_{\rm h}^{~2} - 0.0019 R_{\rm h}^{~3}~ ,                                                              & \mbox{for} & \textit{S9W}, \\
0.956 - 0.0024 R_{\rm h} + 0.007 R_{\rm h}^{~2} - 0.0003 R_{\rm h}^{~3}~ ,                                                              & \mbox{for} & \textit{L18W}, \\
\end{array} \right. 
\label{eq:Ecc}
\end{equation}
where $R_{\rm h}$ is the heliocentric distance in astronomical units (AU).
Given that the filters in the IRC slow-scan observations are the same as those used in the all-sky survey observations, the color correction factors used in our study are the same as those used in the all-sky survey observations \citep{Usui2011}.

\subsection{Geometric Albedo and Diameter Determinations by Thermal Modeling}
In order to analyze a large amount of thermal infrared asteroid data, a simple thermal model such as the standard thermal model (STM; \cite{Lebofsky1986}), isothermal latitude model (ILM; \cite{Veeder1989}), or near-Earth asteroid thermal model (NEATM; \cite{Harris1998}) is generally applied \citep{Harris2002}.
The STM is a simple empirical model that assumes an asteroid is a non-rotating spherical object with zero thermal inertia; consequently, there is no emission on the night side.
The thermal emission from a point on the surface of an asteroid is instantaneously in equilibrium with the solar flux absorbed at that point.
The ILM is also known as the fast rotating model, and is regarded as a model that makes assumptions that are the opposite to those of the STM.
The ILM considers that the surface temperature distribution depends only on latitude, and that the day and night sides contribute equal emissions.
The ILM is of most relevance to an asteroid with a high thermal inertia and/or fast rotation.
The NEATM model is a modified STM that accounts for cases intermediate between zero (STM) and high thermal inertia (ILM).

In our study, the NEATM was used to determine the geometric albedos and diameters of the observed asteroids.
The surface temperature $T(\theta, \varphi)$ is expressed by:

\begin{equation} 
T(\theta, \varphi) = \left\{ \begin{array}{lcl}
T_{\rm SS} \cdot \sin^{1/4} \theta \cdot \cos^{1/4} \varphi~ , & \mbox{for} & 0 \leq \theta \leq {\pi}, \\
0~ ,                                                              & \mbox{for} & {\pi} \leq \theta \leq 2{\pi}, \\
\end{array} \right. 
\label{eq:temperature}
\end{equation}
where $\theta$ and $\varphi$ are the longitude and latitude of the asteroid, respectively. 
The temperature at the sub-solar point $T_{\rm SS}$ in units of kelvin is given by:

\begin{eqnarray}
T_{\rm SS} &=& \left(\frac{(1-A_{\rm B})S_{\rm s}}{\eta \varepsilon \sigma R_{\rm h}^{~2}}\right)^{1/4}~ , 
\label{eq:max temperature}
\end{eqnarray}
where $S_{\rm s}$ is the incident solar flux at the heliocentric distance $R_{\rm h}$ = 1 [AU], $\varepsilon$ denotes the emissivity at infrared wavelengths (we assume a value of 0.9), $\sigma$ is the Stefan--Boltzmann constant, and \textit{$\eta$} is the beaming parameter that accounts for the physical quantities related to the surface roughness and thermal inertia of the asteroid.
The bolometric albedo $A_{\rm B}$ is given by the equation:

\begin{eqnarray}
A_{\rm B} = p_{\rm v} q~ ,
\label{eq:bond albedo}
\end{eqnarray}
where $p_{\rm v}$ and $q$ are the geometric albedo (defined as the ratio of the brightness of an object observed at a zero phase angle to that of a perfectly diffusing Lambertian disk of the same radius located at the same distance) and the phase integral is given by the standard $H$--$G$ system \citep{Bowell1989}:

\begin{eqnarray}
q = 0.290 + 0.684 G~ ,
\label{eq:phase integral}
\end{eqnarray}
where $G$ is the slope parameter. 
The absolute visual magnitude $H$ is obtained from optical photometry, and allows a diameter $D$ in units of kilometers to be estimated for a given albedo from the following expression (e.g., \cite{Pravec2007}): 

\begin{eqnarray}
D &=& \frac{1329}{\sqrt{p_{\rm v}}}~10^{-H/5} ,
\label{eq:relation of d and H}
\end{eqnarray}
Given that the NEATM is defined at zero phase, a thermal infrared phase coefficient is applied as a function of 0.01 mag $\mathrm{deg^{-1}}$ for the NEATM.

Geometry is determined by the heliocentric distance $R_{\rm h}$, the AKARI-centric distance $\Delta$, and the phase angle $\alpha$ used in asteroid identification. 
The absolute magnitude \textit{H} and slope parameter \textit{G} were taken from the dataset of the Lowell Observatory, as was the case for the orbital elements.
For error calculations, we assign MPCORB values of Table 3 in \citet{Pravec2012} as uncertainties of for \textit{H} and 0.10 from \citet{Pravec2012} as it of for \textit{G}.

The NEATM incorporates a beaming parameter \textit{$\eta$}, which takes into account physical quantities related to the surface roughness and thermal inertia of an asteroid.
In the STM, the beaming parameter was held constant at \textit{$\eta$} = 0.756, based on ground-truth occultation observations of the calibrator asteroids.
In the ILM model, \textit{$\eta$} is equal to $\pi$. 
As {AKARI} cannot conduct observations simultaneously in \textit{S9W} and \textit{L18W} bands, default values of \textit{$\eta$} were used. 

For the AcuA dataset, \textit{$\eta$} for the 9 \micron~ data and \textit{$\eta$} for the 18 \micron~ data were used, which provide the best match for the sizes and albedos of 55 large main-belt asteroids \citep{Usui2011}.
These 55 asteroidal calibrators are distributed between 70 and 1,000 km.
Conversely, WISE observed main-belt asteroids between 1 and 1,000 km \citep{Masiero2011}.
For main-belt asteroids smaller than 20 km, beaming values are 1.01 $\pm$ 0.16.
\citet{Masiero2011} used \textit{$\eta$} = 1.0 for asteroids with only a single thermal band as a reasonable assumption for objects in the main-belt.
\citet{Delbo2003} also suggested that default values of \textit{$\eta$} = 1.0 for $\alpha$ $<$  \timeform{30D} be used for the NEATM.
The size range of asteroids observed by \citet{Delbo2003} under $\alpha$ $<$  \timeform{30D} is 0.1--10 km.
Furthermore, \citet{Delbo2007} reported that thermal inertia increases with decreasing size.
\citet{Hasegawa2002} and \citet{Ryan2010} concluded that asteroids with smaller diameters have a much greater range of \textit{$\eta$}.
These trend in thermal inertia and $\eta$ can be explained by the differing regolith properties of small and large asteroids.
The diameter and phase angle of observed asteroids in the IRC slow-scan observations is mainly in the range of 2--20 km and $\alpha$ $\leq$ \timeform{30D}, respectively.
Therefore, all the observed asteroids are assigned \textit{$\eta$} = 1.0 in our study.

\section{Results and Discussion}
In total, 89 sightings of 88 different asteroids were identified from IRC slow-scan observational images.
Table \ref{tab:Observational circumstances} presents relevant observational details and physical data for the asteroids.
The results of the objects included in the catalog are named the Asteroid catalog using AKARI IRC Slow-Scan observations (AcuA-ISS).
The asteroid (51250) 2000 $\mathrm{JO_{47}}$ was only observed in different epochs at each band (2007/07/04 16:17:06 for \textit{L18W} and 2007/07/05 00:34:24 for \textit{S9W}).
All the observed asteroids belong to the main-belt, and include two Hilda asteroids: (13381) 1998 $\mathrm{WJ_{17}}$ and (39301) 2001 $\mathrm{OB_{100}}$.
Results from the IRC slow-scan observations are given in Table \ref{tab:result}, which lists 88 asteroids detected in the mid-infrared region along with their size and albedo data, and associated uncertainties.

\begin{figure}
  \begin{center}
    \FigureFile(80mm,80mm){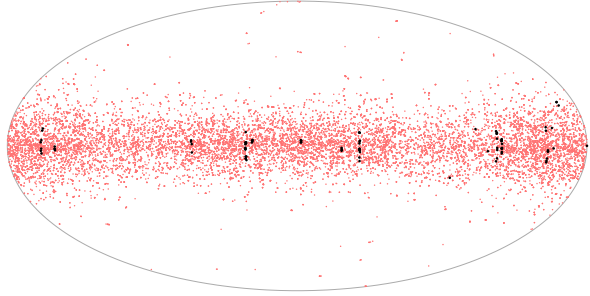}
  \end{center}
  \caption{Distribution of asteroid sightings in ecliptic coordinates. The figure is plotted as a Hammer--Aitoff projection.
The red and black dots indicate the asteroids detected by AucA and AcuA-ISS, respectively.
}
\label{fig:sightings}
\end{figure}

Figure \ref{fig:sightings} shows the asteroids observed by {AKARI}, with asteroids detected from the all-sky survey \citep{Usui2011} highlighted in red and those from the IRC slow-scan observations (this study) in black.
The 16.32 $\mathrm{deg^2}$ area in the ecliptic plane ({$|\beta|$} $\leq$ \timeform{30D}) was observed with the IRC slow-scan observation mode in both the \textit{S9W} and \textit{L18W} filter bands.

\begin{longtable}{rlllccccc}
  \caption{Observational details of the IRC slow-scan mode}\label{tab:Observational circumstances}
  \hline              
  NUM & NAME & PROV & Date & $R_{\rm h}$ [AU] & $\Delta$ [AU] & $\alpha$ [$\timeform{D}$] & PID & MODE
\\ 
\endfirsthead
  \hline
  NUM & NAME & PROV & Date & $R_{\rm h}$ [AU] & $\Delta$ [AU] & $\alpha$ [$\timeform{D}$] & PID & MODE
\\
  \hline
\endhead
  \hline
\endfoot
  \hline
\multicolumn{1}{@{}l}{\rlap{\parbox[t]{1.0\textwidth}{\small
   Notes. 
NUM, NAME, and PROV in the table refer to the number, name, and provisional designation of the asteroid, respectively.
$R_{\rm h}$, $\Delta$, and $\alpha$ are the heliocentric distance, the AKARI-centric distance, and the phase angle, respectively.
PID is the pointed ID for {AKARI}.
Usually it is identical with the Target ID, but is different for parallel mode observations.
MODE indicates the AOT and filter band. 
"-S" and "-L" in MODE mean the \textit{S9W} and the \textit{L18W} bands, respectively.
}}}
\endlastfoot
  \hline
43 &  Ariadne  &    & 2006.11.17 & 2.538602 & 2.334816 & 23.16 & 1500505 &  IRC11-L\\
120 &  Lachesis  &    & 2007.8.25 & 3.297066 & 3.147807 & 17.55 & 1501528 &  IRC51-S\\
303 &  Josephina  &    & 2007.5.27 & 3.227159 & 3.074105 & 18.15 & 1501507 &  IRC51-L\\
389 &  Industria  &  1894 BB  & 2007.2.19 & 2.640165 & 2.449639 & 22.33 & 1500560 &  IRC51-L\\
787 &  Moskva  &  1914 UQ  & 2007.1.19 & 2.678289 & 2.490071 & 21.34 & 1500536 &  IRC51-L\\
1223 &  Neckar  &  1931 TG  & 2007.1.19 & 2.993022 & 2.818871 & 19.04 & 1500536 &  IRC51-S\\
1671 &  Chaika  &  1934 TD  & 2007.7.19 & 3.167706 & 3.010767 & 18.77 & 1501520 &  IRC51-S\\
2727 &  Paton  &  1979 SO9  & 2007.2.18 & 2.611124 & 2.417393 & 21.98 & 1500552 &  IRC51-S\\
2878 &  Panacea  &  1980 RX  & 2007.7.17 & 2.826768 & 2.633587 & 20.44 & 1501516 &  IRC51-S\\
2974 &  Holden  &  1955 QK  & 2007.7.19 & 2.410074 & 2.194508 & 25.12 & 1501520 &  IRC51-S\\
3380 &  Awaji  &  1940 EF  & 2006.12.19 & 2.799506 & 2.627247 & 20.75 & 1500520 &  IRC11-L\\
3454 &  Lieske  &  1981 $\mathrm{WB_{1}}$  & 2007.2.18 & 2.506598 & 2.310919 & 22.99 & 1500552 &  IRC51-L\\
3525 &  Paul  &  1983 $\mathrm{CX_{2}}$  & 2007.5.25 & 3.350142 & 3.209716 & 17.71 & 1501502 &  IRC51-L\\
4508 &  Takatsuki  &  1990 $\mathrm{FG_{1}}$  & 2007.5.25 & 2.107394 & 1.856114 & 29.25 & 1501501 &  IRC51-S\\
5031 &  Svejcar  &  1990 $\mathrm{FW_{1}}$  & 2007.8.25 & 2.734255 & 2.557551 & 21.47 & 1501525 &  IRC51-L\\
5414 &  Sokolov  &  1977 $\mathrm{RW_{6}}$  & 2006.11.18 & 2.654776 & 2.478728 & 21.47 & 1500513 &  IRC11-L\\
5921 &    &  1992 UL  & 2007.5.27 & 2.101870 & 1.853744 & 28.56 & 1501507 &  IRC51-L\\
6113 &  Tsap  &  1982 $\mathrm{SX_{5}}$  & 2007.1.19 & 2.831880 & 2.652805 & 20.01 & 1500537 &  IRC51-L\\
7279 &  Hagfors  &  1985 $\mathrm{VD_{1}}$  & 2007.7.17 & 2.589649 & 2.391503 & 22.91 & 1501513 &  IRC51-L\\
7514 &    &  1986 ED  & 2007.2.19 & 2.659431 & 2.470582 & 22.14 & 1500560 &  IRC51-L\\
7859 &  Lhasa  &  1979 US  & 2006.11.18 & 2.522808 & 2.330671 & 22.63 & 1500513 &  IRC11-S\\
8040 &  Utsumikazuhiko  &  1993 $\mathrm{SY_{3}}$  & 2006.11.16 & 2.387672 & 2.188442 & 24.50 & 1500506 &  IRC11-L\\
8801 &    &  1981 $\mathrm{EQ_{29}}$  & 2007.2.19 & 3.046047 & 2.882508 & 19.11 & 1500561 &  IRC51-L\\
9429 &  Porec  &  1996 $\mathrm{EW_{1}}$  & 2007.1.19 & 3.251622 & 3.097513 & 17.39 & 1500537 &  IRC51-L\\
10224 &  Hisashi  &  1997 $\mathrm{UK_{22}}$  & 2006.12.19 & 2.790197 & 2.615382 & 20.77 & 1500520 &  IRC11-L\\
10442 &  Biezenzo  &  4062 T-1  & 2007.5.25 & 3.237080 & 3.087467 & 18.18 & 1501504 &  IRC51-L\\
10906 &    &  1997 $\mathrm{WO_{44}}$  & 2007.1.18 & 2.065941 & 1.824184 & 27.53 & 1500538 &  IRC51-S\\
11296 &  Denzen  &  1992 KA  & 2007.5.27 & 2.680008 & 2.491249 & 22.08 & 1501507 &  IRC51-L\\
12206 &    &  1981 $\mathrm{EG_{27}}$  & 2007.1.19 & 2.845371 & 2.667759 & 19.91 & 1500537 &  IRC51-L\\
12334 &    &  1992 $\mathrm{WD_{3}}$  & 2006.11.17 & 2.032610 & 1.767803 & 29.39 & 1500505 &  IRC11-S\\
12411 &  Tannokayo  &  1995 $\mathrm{SQ_{3}}$  & 2007.2.19 & 2.146342 & 1.908209 & 27.86 & 1500560 &  IRC51-L\\
13149 &  Heisenberg  &  1995 $\mathrm{EF_{8}}$  & 2007.2.18 & 3.298470 & 3.153460 & 17.31 & 1500552 &  IRC51-L\\
13381 &    &  1998 $\mathrm{WJ_{17}}$  & 2007.2.19 & 4.215680 & 4.098958 & 13.65 & 1500561 &  IRC51-L\\
13445 &    &  3063 P-L  & 2007.1.18 & 3.298641 & 3.163087 & 16.94 & 1500538 &  IRC51-L\\
14147 &  Wenlingshuguang  &  1998 $\mathrm{SG_{43}}$  & 2006.11.17 & 2.341921 & 2.120256 & 25.50 & 1500504 &  IRC11-L\\
15746 &    &  1991 $\mathrm{PN_{8}}$  & 2006.11.17 & 2.670670 & 2.477771 & 21.97 & 1500505 &  IRC11-L\\
16154 &  Dabramo  &  2000 $\mathrm{AW_{2}}$  & 2007.1.18 & 3.321473 & 3.186623 & 16.86 & 1500538 &  IRC51-L\\
23916 &    &  1998 $\mathrm{SD_{131}}$  & 2007.2.18 & 3.056860 & 2.891344 & 18.57 & 1500553 &  IRC51-S\\
24389 &    &  2000 $\mathrm{AA_{177}}$  & 2007.5.25 & 2.429128 & 2.218058 & 24.81 & 1501502 &  IRC51-S\\
29027 &    &  7587 P-L  & 2007.2.19 & 2.290763 & 2.062212 & 26.01 & 1500560 &  IRC51-S\\
29352 &    &  1995 JR  & 2007.7.19 & 2.530867 & 2.333805 & 23.79 & 1501520 &  IRC51-L\\
29664 &    &  1998 $\mathrm{WY_{23}}$  & 2007.2.19 & 2.738363 & 2.548114 & 21.38 & 1500561 &  IRC51-S\\
31062 &    &  1996 $\mathrm{TP_{10}}$  & 2007.8.25 & 3.127006 & 2.965808 & 18.78 & 1501527 &  IRC51-S\\
34218 &    &  2000 $\mathrm{QC_{78}}$  & 2007.1.20 & 2.640249 & 2.461623 & 21.87 & 1500546 &  IRC51-L\\
34301 &    &  2000 $\mathrm{QO_{171}}$  & 2006.11.17 & 2.731269 & 2.545068 & 21.57 & 1500504 &  IRC11-L\\
34676 &    &  2000 $\mathrm{YF_{126}}$  & 2006.9.21 & 3.453976 & 3.315299 & 16.91 & 1710038 &  IRC51-L\\
37345 &    &  2001 $\mathrm{SV_{153}}$  & 2007.8.25 & 3.377545 & 3.241115 & 17.19 & 1501526 &  IRC51-L\\
38120 &    &  1999 $\mathrm{JN_{39}}$  & 2007.5.25 & 1.923790 & 1.645326 & 32.59 & 1501501 &  IRC51-S\\
39301 &    &  2001 $\mathrm{OB_{100}}$  & 2007.7.19 & 4.500299 & 4.400443 & 13.07 & 1501520 &  IRC51-L\\
44191 &    &  1998 $\mathrm{LF_{2}}$  & 2007.1.20 & 2.625713 & 2.446439 & 21.92 & 1500546 &  IRC51-L\\
46072 &    &  2001 EJ  & 2007.1.19 & 2.959495 & 2.783794 & 19.28 & 1500536 &  IRC51-S\\
48571 &    &  1994 $\mathrm{ER_{5}}$  & 2007.2.19 & 2.327067 & 2.108739 & 25.59 & 1500560 &  IRC51-L\\
49402 &    &  1998 $\mathrm{XZ_{44}}$  & 2007.8.25 & 2.528788 & 2.331854 & 23.08 & 1501528 &  IRC51-L\\
51250 &    &  2000 $\mathrm{JO_{47}}$  & 2007.7.4 & 2.918787 & 2.749098 & 19.32 & 1602431 &  IRC51-S\\
51250 &    &  2000 $\mathrm{JO_{47}}$  & 2007.7.5 & 2.919128 & 2.745111 & 19.30 & 1602428 &  IRC51-L\\
52265 &    &  1985 $\mathrm{RM_{3}}$  & 2007.1.19 & 2.508926 & 2.306929 & 22.90 & 1500536 &  IRC51-L\\
55883 &    &  1997 $\mathrm{WF_{8}}$  & 2006.11.16 & 2.389337 & 2.186822 & 25.17 & 1500503 &  IRC11-S\\
57208 &    &  2001 $\mathrm{QB_{57}}$  & 2007.1.19 & 2.272077 & 2.049657 & 25.85 & 1500545 &  IRC51-L\\
57481 &    &  2001 $\mathrm{ST_{153}}$  & 2006.12.19 & 2.288468 & 2.067001 & 25.66 & 1500520 &  IRC11-S\\
58011 &    &  2002 $\mathrm{TW_{280}}$  & 2007.5.25 & 3.001184 & 2.837739 & 19.63 & 1501504 &  IRC51-L\\
58748 &    &  1998 $\mathrm{FB_{9}}$  & 2007.1.19 & 2.718304 & 2.535665 & 21.36 & 1500545 &  IRC51-L\\
58798 &    &  1998 $\mathrm{FU_{100}}$  & 2007.2.19 & 3.116269 & 2.956199 & 18.63 & 1500561 &  IRC51-L\\
63717 &    &  2001 $\mathrm{QT_{220}}$  & 2007.2.19 & 2.203776 & 1.971396 & 26.89 & 1500561 &  IRC51-L\\
63887 &    &  2001 $\mathrm{SH_{3}}$  & 2007.8.25 & 3.207130 & 3.051977 & 18.06 & 1501528 &  IRC51-S\\
74318 &    &  1998 $\mathrm{UB_{16}}$  & 2007.8.25 & 3.001487 & 2.842352 & 19.51 & 1501525 &  IRC51-L\\
79087 &  Scheidt  &  1977 $\mathrm{UM_{2}}$  & 2007.1.20 & 2.167639 & 1.945359 & 27.89 & 1500543 &  IRC51-L\\
81955 &    &  2000 $\mathrm{PT_{18}}$  & 2007.1.19 & 2.943009 & 2.774224 & 19.67 & 1500545 &  IRC51-L\\
85871 &    &  1999 $\mathrm{BN_{30}}$  & 2007.2.28 & 2.298378 & 2.082549 & 25.72 & 1600365 &  IRC51-L\\
93418 &    &  2000 $\mathrm{SB_{305}}$  & 2007.1.18 & 2.619743 & 2.441709 & 21.43 & 1500538 &  IRC51-L\\
93798 &    &  2000 $\mathrm{WP_{45}}$  & 2007.1.18 & 2.821893 & 2.660610 & 20.49 & 1500535 &  IRC51-L\\
94193 &    &  2001 $\mathrm{BN_{7}}$  & 2007.5.27 & 3.066873 & 2.905949 & 19.16 & 1501507 &  IRC51-L\\
109898 &    &  2001 $\mathrm{SE_{20}}$  & 2006.11.18 & 3.305919 & 3.169253 & 17.12 & 1500513 &  IRC11-L\\
114424 &    &  2002 $\mathrm{YE_{36}}$  & 2006.11.16 & 2.163698 & 1.942199 & 28.08 & 1500503 &  IRC11-L\\
116491 &    &  2004 $\mathrm{BU_{12}}$  & 2007.5.25 & 3.096796 & 2.940222 & 19.35 & 1501501 &  IRC51-L\\
119811 &    &  2002 $\mathrm{AC_{156}}$  & 2006.11.17 & 2.746422 & 2.558025 & 21.29 & 1500505 &  IRC11-L\\
130992 &    &  2000 $\mathrm{WT_{159}}$  & 2007.8.25 & 2.344822 & 2.128820 & 25.45 & 1501527 &  IRC51-L\\
131490 &    &  2001 $\mathrm{SJ_{164}}$  & 2007.8.25 & 2.762148 & 2.585283 & 21.03 & 1501528 &  IRC51-L\\
135859 &    &  2002 TS  & 2006.11.18 & 2.387461 & 2.186881 & 24.04 & 1500513 &  IRC11-L\\
139668 &    &  2001 $\mathrm{QB_{194}}$  & 2006.11.17 & 2.318870 & 2.094051 & 25.56 & 1500505 &  IRC11-L\\
140692 &    &  2001 $\mathrm{UF_{65}}$  & 2007.1.20 & 2.676858 & 2.497344 & 22.09 & 1500543 &  IRC51-S\\
141308 &    &  2001 $\mathrm{YT_{119}}$  & 2007.1.20 & 2.910120 & 2.750073 & 19.72 & 1500546 &  IRC51-L\\
147957 &    &  1993 $\mathrm{TM_{21}}$  & 2006.11.16 & 2.147420 & 1.918652 & 28.33 & 1500503 &  IRC11-S\\
150620 &    &  2000 $\mathrm{YM_{71}}$  & 2006.11.16 & 2.550888 & 2.361453 & 22.75 & 1500506 &  IRC11-S\\
150739 &    &  2001 $\mathrm{QA_{67}}$  & 2006.12.19 & 2.412085 & 2.202069 & 24.31 & 1500520 &  IRC11-S\\
155875 &    &  2001 $\mathrm{DR_{91}}$  & 2007.1.19 & 2.984466 & 2.814412 & 18.98 & 1500537 &  IRC51-L\\
157836 &    &  1998 $\mathrm{FG_{102}}$  & 2007.1.18 & 2.249671 & 2.031143 & 25.02 & 1500538 &  IRC51-S\\
162143 &    &  1998 $\mathrm{VJ_{1}}$  & 2007.7.4 & 2.065237 & 1.804591 & 27.61 & 1602429 &  IRC51-S\\
249789 &    &  2000 $\mathrm{XC_{53}}$  & 2006.11.17 & 2.565457 & 2.360142 & 23.14 & 1500504 &  IRC11-S\\
253116 &    &  2002 $\mathrm{UX_{69}}$  & 2007.3.7 & 1.919815 & 1.648152 & 31.59 & 3160010 &  IRC51-S\\
\end{longtable}

\begin{longtable}{rllrcrcccccc}
  \caption{Geometric albedo and diameter data for asteroids detected by IRC slow-scan observations}\label{tab:result}
  \hline              
  NUM & NAME & PROV & \textit{$H$} [mag] & \textit{$G$} & \textit{$D$} [km] & $\sigma_{D}$ [km] & \textit{$p_{\rm v}$} & $\sigma_{p_{\rm v}}$ & S & A & W
\\ 
\endfirsthead
  \hline
  NUM & NAME & PROV & \textit{$H$} [mag] & \textit{$G$} & \textit{$D$} [km] & $\sigma_{D}$ [km] & \textit{$p_{\rm v}$} & $\sigma_{p_{\rm v}}$ & S & A & W
\\
  \hline
\endhead
  \hline
\endfoot
  \hline
\multicolumn{1}{@{}l}{\rlap{\parbox[t]{1.0\textwidth}{\small
   Notes. 
\textit{$H$} and \textit{$G$} are the absolute magnitude and slope parameter taken from the Asteroid Orbital Elements Database of the Lowell Observatory, respectively.
\textit{$D$} and \textit{$p_{\rm v}$} are the calculated diameter and geometric albedo of the asteroids, respectively.
$\sigma_{D}$ and $\sigma_{p_{\rm v}}$ are the uncertainties on \textit{$D$} and \textit{$p_{\rm v}$}, respectively, estimated from the thermal model calculations.
S, A, and W indicate whether the asteroid was detected by {IRAS} \citep{Tedesco2002b}, {AKARI} \citep{Usui2011}, or {WISE} (\cite{Masiero2011}, \cite{Grav2012}), respectively.
}}}
\endlastfoot
  \hline
  43 & Ariadne &  &  7.93 & 0.11 & 57.56 & 1.68 & 0.359 & 0.040 & 1 & 1 & 1\\
  120 & Lachesis &  &  7.75 & 0.15 & 197.13 & 2.74 & 0.036 & 0.004 & 1 & 1 & 1\\
  303 & Josephina &  &  8.80 & 0.15 & 105.36 & 0.54 & 0.048 & 0.007 & 1 & 1 & 1\\
  389 & Industria & 1894 BB &  7.88 & 0.15 & 89.15 & 0.28 & 0.157 & 0.015 & 1 & 1 & 0\\
  787 & Moskva & 1914 UQ &  9.70 & 0.15 & 29.42 & 0.28 & 0.269 & 0.040 & 1 & 1 & 1\\
  1223 & Neckar & 1931 TG &  10.58 & 0.15 & 26.07 & 0.86 & 0.152 & 0.025 & 0 & 1 & 1\\
  1671 & Chaika & 1934 TD &  12.10 & 0.15 & 13.29 & 1.71 & 0.145 & 0.043 & 0 & 0 & 1\\
  2727 & Paton & 1979 SO9 &  12.10 & 0.15 & 8.92 & 0.88 & 0.321 & 0.080 & 0 & 0 & 1\\
  2878 & Panacea & 1980 RX &  11.70 & 0.15 & 17.00 & 0.76 & 0.128 & 0.022 & 0 & 0 & 1\\
  2974 & Holden & 1955 QK &  13.10 & 0.15 & 6.15 & 0.75 & 0.269 & 0.082 & 0 & 0 & 1\\
  3380 & Awaji & 1940 EF &  12.00 & 0.15 & 8.35 & 0.92 & 0.402 & 0.107 & 0 & 0 & 0\\
  3454 & Lieske & 1981 $\mathrm{WB_{1}}$ &  13.20 & 0.15 & 7.13 & 0.38 & 0.183 & 0.039 & 0 & 0 & 1\\
  3525 & Paul & 1983 $\mathrm{CX_{2}}$ &  12.10 & 0.15 & 16.64 & 0.59 & 0.092 & 0.015 & 0 & 1 & 0\\
  4508 & Takatsuki & 1990 $\mathrm{FG_{1}}$ &  13.50 & 0.15 & 3.54 & 0.64 & 0.560 & 0.229 & 0 & 0 & 1\\
  5031 & Svejcar & 1990 $\mathrm{FW_{1}}$ &  14.00 & 0.15 & 8.92 & 0.54 & 0.056 & 0.014 & 0 & 0 & 1\\
  5414 & Sokolov & 1977 $\mathrm{RW_{6}}$ &  12.70 & 0.15 & 6.90 & 1.03 & 0.309 & 0.109 & 0 & 0 & 1\\
  5921 &  & 1992 UL &  13.60 & 0.15 & 4.11 & 0.31 & 0.379 & 0.090 & 0 & 0 & 1\\
  6113 & Tsap & 1982 $\mathrm{SX_{5}}$ &  13.10 & 0.15 & 13.40 & 0.36 & 0.057 & 0.011 & 0 & 1 & 1\\
  7279 & Hagfors & 1985 $\mathrm{VD_{1}}$ &  12.90 & 0.15 & 12.67 & 0.29 & 0.076 & 0.014 & 0 & 0 & 1\\
  7514 &  & 1986 ED &  13.50 & 0.15 & 9.82 & 0.44 & 0.073 & 0.015 & 0 & 0 & 0\\
  7859 & Lhasa & 1979 US &  13.30 & 0.15 & 12.66 & 0.65 & 0.053 & 0.011 & 1 & 0 & 0\\
  8040 & Utsumikazuhiko & 1993 $\mathrm{SY_{3}}$ &  13.70 & 0.15 & 5.14 & 0.64 & 0.221 & 0.069 & 0 & 0 & 0\\
  8801 &  & 1981 $\mathrm{EQ_{29}}$ &  13.30 & 0.15 & 5.35 & 1.35 & 0.296 & 0.159 & 0 & 0 & 1\\
  9429 & Porec & 1996 $\mathrm{EW_{1}}$ &  13.30 & 0.15 & 11.61 & 0.73 & 0.063 & 0.014 & 0 & 0 & 1\\
  10224 & Hisashi & 1997 $\mathrm{UK_{22}}$ &  14.10 & 0.15 & 6.14 & 1.36 & 0.107 & 0.053 & 0 & 0 & 1\\
  10442 & Biezenzo & 4062 T-1 &  12.90 & 0.15 & 15.57 & 0.54 & 0.050 & 0.010 & 0 & 0 & 1\\
  10906 &  & 1997 $\mathrm{WO_{44}}$ &  13.70 & 0.15 & 3.77 & 0.64 & 0.412 & 0.159 & 0 & 0 & 1\\
  11296 & Denzen & 1992 KA &  13.20 & 0.15 & 6.43 & 0.68 & 0.224 & 0.063 & 0 & 0 & 1\\
  12206 &  & 1981 $\mathrm{EG_{27}}$ &  14.70 & 0.15 & 6.81 & 0.59 & 0.050 & 0.014 & 0 & 0 & 0\\
  12334 &  & 1992 $\mathrm{WD_{3}}$ &  13.80 & 0.15 & 6.14 & 0.34 & 0.142 & 0.035 & 0 & 0 & 1\\
  12411 & Tannokayo & 1995 $\mathrm{SQ_{3}}$ &  13.70 & 0.15 & 3.29 & 0.52 & 0.539 & 0.197 & 0 & 0 & 0\\
  13149 & Heisenberg & 1995 $\mathrm{EF_{8}}$ &  13.70 & 0.15 & 5.29 & 1.62 & 0.209 & 0.134 & 0 & 0 & 0\\
  13381 &  & 1998 $\mathrm{WJ_{17}}$ &  12.20 & 0.15 & 18.98 & 1.29 & 0.065 & 0.015 & 0 & 0 & 1\\
  13445 &  & 3063 P-L &  12.60 & 0.15 & 15.94 & 0.56 & 0.063 & 0.012 & 0 & 0 & 0\\
  14147 & Wenlingshuguang & 1998 $\mathrm{SG_{43}}$ &  13.70 & 0.15 & 4.02 & 0.95 & 0.363 & 0.184 & 0 & 0 & 1\\
  15746 &  & 1991 $\mathrm{PN_{8}}$ &  14.40 & 0.15 & 5.54 & 0.70 & 0.100 & 0.034 & 0 & 0 & 0\\
  16154 & Dabramo & 2000 $\mathrm{AW_{2}}$ &  12.30 & 0.15 & 10.41 & 0.83 & 0.196 & 0.048 & 0 & 0 & 0\\
  23916 &  & 1998 $\mathrm{SD_{131}}$ &  13.40 & 0.15 & 15.00 & 1.30 & 0.034 & 0.009 & 0 & 0 & 0\\
  24389 &  & 2000 $\mathrm{AA_{177}}$ &  14.20 & 0.15 & 7.25 & 0.61 & 0.070 & 0.019 & 0 & 0 & 1\\
  29027 &  & 7587 P-L &  14.90 & 0.15 & 3.11 & 1.33 & 0.200 & 0.177 & 0 & 0 & 0\\
  29352 &  & 1995 JR &  14.40 & 0.15 & 5.95 & 0.58 & 0.087 & 0.026 & 0 & 0 & 0\\
  29664 &  & 1998 $\mathrm{WY_{23}}$ &  13.10 & 0.15 & 9.11 & 1.14 & 0.122 & 0.038 & 0 & 0 & 1\\
  31062 &  & 1996 $\mathrm{TP_{10}}$ &  12.00 & 0.15 & 17.68 & 1.21 & 0.090 & 0.018 & 0 & 1 & 1\\
  34218 &  & 2000 $\mathrm{QC_{78}}$ &  14.30 & 0.15 & 3.20 & 1.25 & 0.328 & 0.266 & 0 & 0 & 0\\
  34301 &  & 2000 $\mathrm{QO_{171}}$ &  14.10 & 0.15 & 4.86 & 1.35 & 0.172 & 0.103 & 0 & 0 & 0\\
  34676 &  & 2000 $\mathrm{YF_{126}}$ &  13.70 & 0.15 & 11.21 & 0.43 & 0.047 & 0.009 & 0 & 0 & 1\\
  37345 &  & 2001 $\mathrm{SV_{153}}$ &  13.60 & 0.15 & 10.90 & 0.83 & 0.054 & 0.013 & 0 & 0 & 1\\
  38120 &  & 1999 $\mathrm{JN_{39}}$ &  15.40 & 0.15 & 4.35 & 0.34 & 0.065 & 0.018 & 0 & 0 & 1\\
  39301 &  & 2001 $\mathrm{OB_{100}}$ &  12.70 & 0.15 & 16.77 & 1.62 & 0.052 & 0.014 & 0 & 0 & 0\\
  44191 &  & 1998 $\mathrm{LF_{2}}$ &  15.00 & 0.15 & 5.42 & 0.69 & 0.060 & 0.020 & 0 & 0 & 0\\
  46072 &  & 2001 EJ &  13.60 & 0.15 & 8.62 & 2.09 & 0.086 & 0.045 & 0 & 0 & 1\\
  48571 &  & 1994 $\mathrm{ER_{5}}$ &  15.20 & 0.15 & 4.96 & 0.48 & 0.060 & 0.018 & 0 & 0 & 0\\
  49402 &  & 1998 $\mathrm{XZ_{44}}$ &  14.20 & 0.15 & 8.15 & 0.37 & 0.056 & 0.013 & 0 & 0 & 1\\
  51250 &  & 2000 $\mathrm{JO_{47}}$ &  13.30 & 0.15 & 13.02 & 1.08 & 0.050 & 0.012 & 0 & 0 & 1\\
  51250 &  & 2000 $\mathrm{JO_{47}}$ &  13.30 & 0.15 & 12.61 & 0.41 & 0.053 & 0.010 & 0 & 0 & 1\\
  52265 &  & 1985 $\mathrm{RM_{3}}$ &  15.20 & 0.15 & 4.30 & 0.69 & 0.079 & 0.031 & 0 & 0 & 1\\
  55883 &  & 1997 $\mathrm{WF_{8}}$ &  14.60 & 0.15 & 6.11 & 0.77 & 0.068 & 0.023 & 0 & 0 & 0\\
  57208 &  & 2001 $\mathrm{QB_{57}}$ &  14.60 & 0.15 & 4.08 & 0.53 & 0.153 & 0.052 & 0 & 0 & 1\\
  57481 &  & 2001 $\mathrm{ST_{153}}$ &  14.70 & 0.15 & 6.62 & 0.63 & 0.053 & 0.016 & 0 & 0 & 0\\
  58011 &  & 2002 $\mathrm{TW_{280}}$ &  13.80 & 0.15 & 9.25 & 0.62 & 0.062 & 0.016 & 0 & 0 & 1\\
  58748 &  & 1998 $\mathrm{FB_{9}}$ &  14.70 & 0.15 & 5.05 & 0.85 & 0.091 & 0.037 & 0 & 0 & 1\\
  58798 &  & 1998 $\mathrm{FU_{100}}$ &  13.80 & 0.15 & 7.64 & 0.98 & 0.091 & 0.031 & 0 & 0 & 1\\
  63717 &  & 2001 $\mathrm{QT_{220}}$ &  15.00 & 0.15 & 2.63 & 1.09 & 0.256 & 0.220 & 0 & 0 & 1\\
  63887 &  & 2001 $\mathrm{SH_{3}}$ &  13.00 & 0.15 & 14.06 & 1.43 & 0.056 & 0.015 & 0 & 0 & 1\\
  74318 &  & 1998 $\mathrm{UB_{16}}$ &  14.00 & 0.15 & 7.57 & 0.87 & 0.078 & 0.025 & 0 & 0 & 0\\
  79087 & Scheidt & 1977 $\mathrm{UM_{2}}$ &  14.10 & 0.15 & 3.26 & 0.58 & 0.381 & 0.160 & 0 & 0 & 1\\
  81955 &  & 2000 $\mathrm{PT_{18}}$ &  14.70 & 0.15 & 7.14 & 0.82 & 0.046 & 0.015 & 0 & 0 & 1\\
  85871 &  & 1999 $\mathrm{BN_{30}}$ &  15.00 & 0.15 & 3.22 & 0.45 & 0.171 & 0.061 & 0 & 0 & 0\\
  93418 &  & 2000 $\mathrm{SB_{305}}$ &  15.50 & 0.15 & 7.15 & 0.56 & 0.022 & 0.006 & 0 & 0 & 1\\
  93798 &  & 2000 $\mathrm{WP_{45}}$ &  15.20 & 0.15 & 8.65 & 0.65 & 0.020 & 0.005 & 0 & 0 & 1\\
  94193 &  & 2001 $\mathrm{BN_{7}}$ &  14.40 & 0.15 & 6.96 & 0.92 & 0.063 & 0.022 & 0 & 0 & 1\\
  109898 &  & 2001 $\mathrm{SE_{20}}$ &  14.40 & 0.15 & 8.09 & 1.39 & 0.047 & 0.019 & 0 & 0 & 1\\
  114424 &  & 2002 $\mathrm{YE_{36}}$ &  14.80 & 0.15 & 5.59 & 0.38 & 0.068 & 0.018 & 0 & 0 & 0\\
  116491 &  & 2004 $\mathrm{BU_{12}}$ &  14.60 & 0.15 & 9.54 & 0.76 & 0.028 & 0.008 & 0 & 0 & 1\\
  119811 &  & 2002 $\mathrm{AC_{156}}$ &  14.70 & 0.15 & 5.63 & 0.75 & 0.073 & 0.025 & 0 & 0 & 1\\
  130992 &  & 2000 $\mathrm{WT_{159}}$ &  15.90 & 0.15 & 5.25 & 0.41 & 0.028 & 0.008 & 0 & 0 & 1\\
  131490 &  & 2001 $\mathrm{SJ_{164}}$ &  14.50 & 0.15 & 6.81 & 0.69 & 0.060 & 0.018 & 0 & 0 & 1\\
  135859 &  & 2002 TS &  15.70 & 0.15 & 6.20 & 0.67 & 0.024 & 0.007 & 0 & 0 & 1\\
  139668 &  & 2001 $\mathrm{QB_{194}}$ &  14.70 & 0.15 & 7.39 & 0.34 & 0.043 & 0.010 & 0 & 0 & 0\\
  140692 &  & 2001 $\mathrm{UF_{65}}$ &  14.30 & 0.15 & 9.23 & 1.07 & 0.040 & 0.013 & 0 & 0 & 0\\
  141308 &  & 2001 $\mathrm{YT_{119}}$ &  15.00 & 0.15 & 4.23 & 1.38 & 0.099 & 0.068 & 0 & 0 & 0\\
  147957 &  & 1993 $\mathrm{TM_{21}}$ &  15.70 & 0.15 & 2.59 & 1.37 & 0.139 & 0.150 & 0 & 0 & 0\\
  150620 &  & 2000 $\mathrm{YM_{71}}$ &  15.40 & 0.15 & 4.46 & 2.47 & 0.062 & 0.070 & 0 & 0 & 1\\
  150739 &  & 2001 $\mathrm{QA_{67}}$ &  14.70 & 0.15 & 8.73 & 0.62 & 0.031 & 0.008 & 0 & 0 & 0\\
  155875 &  & 2001 $\mathrm{DR_{91}}$ &  15.00 & 0.15 & 9.65 & 0.62 & 0.019 & 0.005 & 0 & 0 & 0\\
  157836 &  & 1998 $\mathrm{FG_{102}}$ &  15.20 & 0.15 & 4.51 & 0.81 & 0.072 & 0.031 & 0 & 0 & 1\\
  162143 &  & 1998 $\mathrm{VJ_{1}}$ &  15.10 & 0.15 & 4.28 & 0.92 & 0.088 & 0.043 & 0 & 0 & 1\\
  249789 &  & 2000 $\mathrm{XC_{53}}$ &  14.60 & 0.15 & 8.78 & 0.97 & 0.033 & 0.010 & 0 & 0 & 0\\
  253116 &  & 2002 $\mathrm{UX_{69}}$ &  16.60 & 0.15 & 2.60 & 0.23 & 0.060 & 0.017 & 0 & 0 & 1\\
\end{longtable}

\begin{figure}
  \begin{center}
    \FigureFile(80mm,80mm){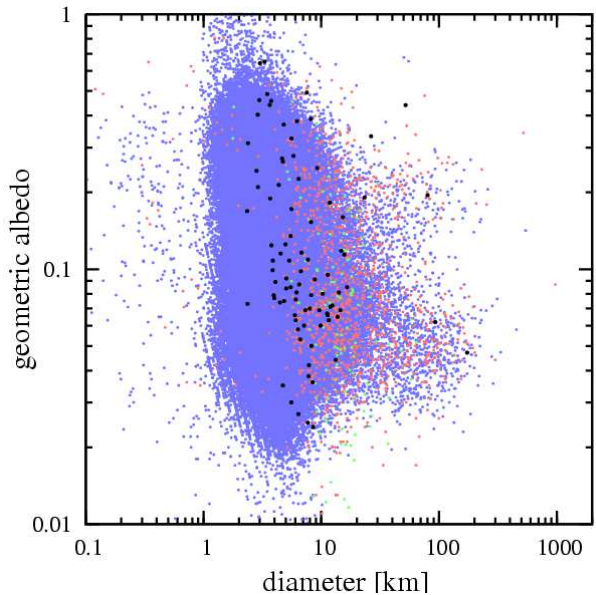}
  \end{center}
  \caption{Distribution of asteroidal diameter and geometric albedo data in the catalogs of SIMPS, AcuA, {WISE}, and this study (AcuA-ISS). 
The green, red, blue, and black dots represent the asteroids listed in the SIMPS, AcuA, {WISE}, and AcuA-ISS catalogs, respectively.
}
\label{fig:distribution}
\end{figure}

Figure \ref{fig:distribution} shows the distribution of albedos as a function of diameter from the asteroid catalogs of SIMPS, AcuA, {WISE}, and this study.
The detection limits of the IRC slow-scan observations are considered to be 9 mJy for the \textit{S9W} band and 30 mJy for the \textit{L18W} \citep{Takita2012}.
The sensitivity of the IRC slow-scan observations is better than that of the {AKARI} all-sky survey \citep{Ishihara2010}, but worse than that of the {WISE} survey \citep{Wright2010}.
The detection limits of asteroids observed by the different surveys are consistent with the relative sensitivity of the surveys. 

A total of 6, 9, and 56 asteroids detected by our slow-scan observations were previously recorded in the SIMPS, AcuA, and {WISE} catalogs, respectively.
There is no overlap in the {ISO} and {MSX} observations (\cite{Mueller2002}, \cite{Dotto2002}, \cite{Tedesco2002a}).
The geometric albedo and diameter of 29 asteroids in the AcuA-ISS appeared first.
All the observed asteroids in the AcuA-ISS catalog should have been originally detected by {WISE}.
However, as {WISE} was not able to investigate all asteroids in the main-belt at mid-infrared wavelengths (e.g., Figure 1 of \citet{Masiero2011}), these 29 asteroids were not actually observed by {WISE}.

\begin{figure}
  \begin{center}
    \FigureFile(80mm,80mm){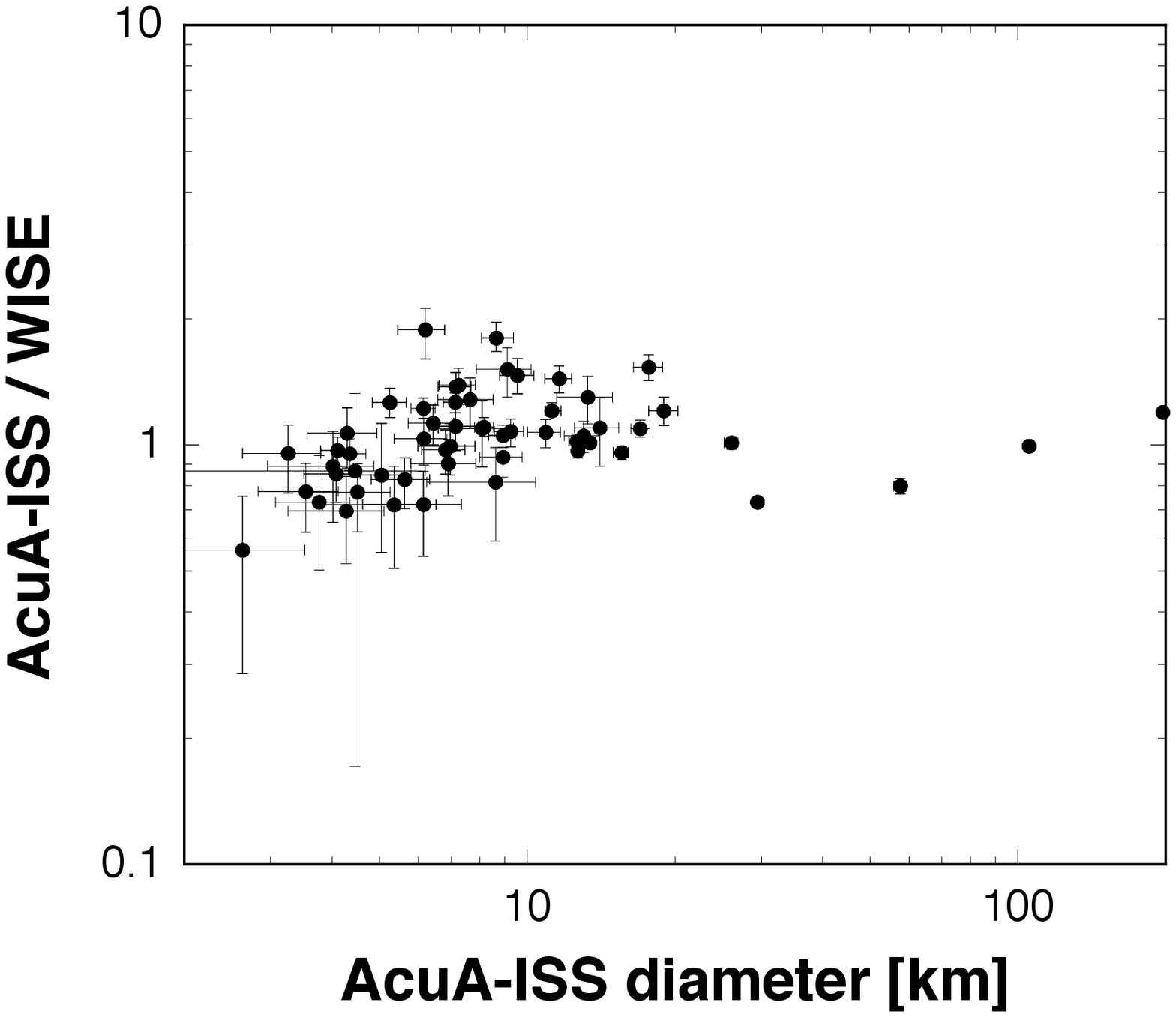}
    \FigureFile(80mm,80mm){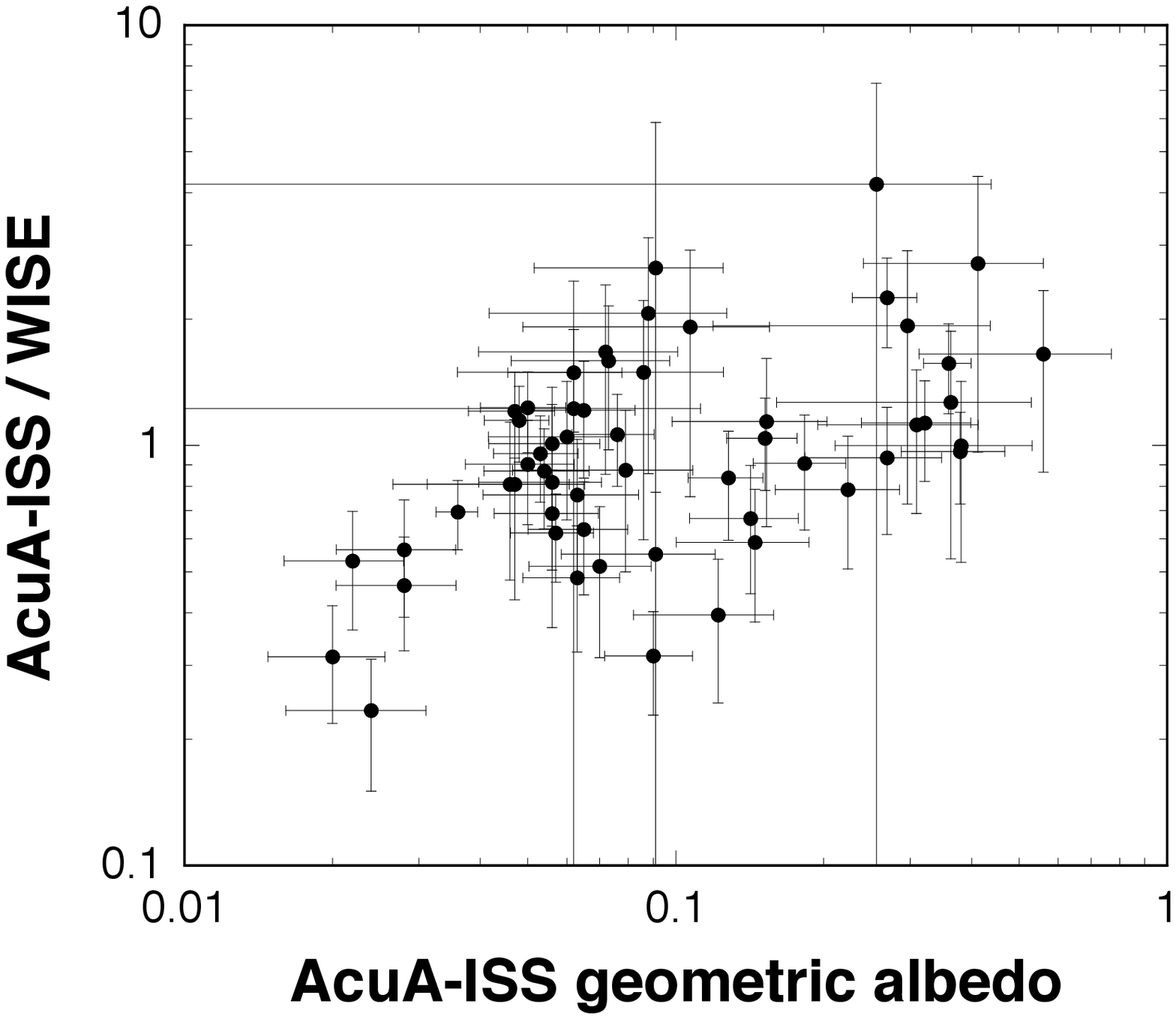}
  \end{center}
  \caption{Comparison between the geometric albedo and diameter data from WISE and the slow-scan observational measurements of this study.}
\label{fig:hikaku}
\end{figure}

Figure \ref{fig:hikaku} compares the differences between the geometric albedo and diameter data obtained by IRC slow-scan observations and {WISE}.
The differences are ca. 6\% for the diameter and ca. 11\% for the geometric albedo.
These differences are comparable to the range of the mean error of the asteroids observed by {WISE} (diameter: 5\%; geometric albedo: 21\%).
Most of the absolute visual magnitudes potentially have large uncertainties, partly due to some hidden light curve effects.
The uncertainty on the absolute visual magnitude does not, however, influence the diameter calculation significantly, even though it changes the derived albedos significantly. 
This accounts for why the differences in albedos are larger than the differences in the diameter data from the various surveys.
However, despite this, the IRC slow-scan observational measurements are in good agreement with previous results.

Given that the peak in the incremental number in the AcuA-ISS is ca. 8 km in diameter, main-belt asteroids in the AcuA-ISS were flux limited to an asteroid diameter threshold of \textgreater{}8 km.
The cumulative number of asteroids \textgreater{}10 km in size in the AcuA-ISS catalog is 24.
{WISE} observed 7090 asteroids \textgreater{}10 km in diameter (\cite{Masiero2011}, \cite{Grav2012}).
The ratio of detected objects with AcuA-ISS and {WISE} is $\sim$ 3 $\times$ $10^{-3}$. 
Conversely, almost all the asteroids in the AcuA-ISS catalog were detected in {$|\beta|$} $\leq$ \timeform{10D}.
The IRC slow-scan observations were performed to obtain 34 images approximately 26.4 $\mathrm{deg^{2}}$ in {$|\beta|$} $\leq$ \timeform{10D}.
As the {WISE} mission life was about 8 months, the size of the survey area in {$|\beta|$} $\leq$ \timeform{10D} for {WISE} is equal to about 9.5 $\times$ $10^{3}$ $\mathrm{deg^{2}}$.
The difference in the observed area between AcuA-ISS and {WISE} is $\sim$ 3 $\times$ $10^{-3}$, which as expected is consistent with the detection ratio of AcuA-ISS and {WISE} given that detection efficiency depends on the search area.

Twenty-six objects in the AcuA-ISS catalog were classified by \citet{Tholen1984}, \citet{Xu1995}, \citet{Bus2002}, \citet{Lazzaro2004}, and \citet{Carvano2010}.
The asteroids (12334) 1992 $\mathrm{WD_{3}}$, (34301) 2000 $\mathrm{QO_{171}}$, and (46072) 2001 EJ belong to the CX spectral type \citep{Carvano2010}.
Albedo values suggest that the two asteroids (34301) 2000 $\mathrm{QO_{171}}$ and (12334) 1992 $\mathrm{WD_{3}}$ are M-type and the asteroid (46072) 2001 EJ is C- or P-type.
The taxonomic type of the asteroid (16154) Dabramo is X-type \citep{Carvano2010}.
However, because the geometric albedo of (16154) Dabramo is \textgreater{}0.1, it is classified as an M-type asteroid.
The geometric albedos of the asteroids (2878) Panacea and (13149) Heisenberg are \textgreater{}0.1, but are classed as D- and C-type asteroids, respectively.
\citet{Mainzer2012} and Usui et al. (ApJ, submitted) indicated that C- and D-type asteroids having high albedo are present from \textless{}50 km in size.
Therefore, these asteroids are rare C or D-type asteroids distinguished by high albedos.

Twenty-seven asteroids in the AcuA-ISS catalog belong to the Vesta family \citep{Nesvorny2006}.
The spectral type of the two asteroids (2974) Holden and (63717) 2001 $\mathrm{QT_{220}}$ are unknown.
The geometric albedos of these asteroids are \textgreater{}0.25.
Taking into account the high albedo and link to the Vesta family, these asteroids may be V-type.

\section{Summary}
A serendipitous asteroidal catalog (AcuA-ISS; 88 objects) has been established from slow-scan observations at mid-infrared wavelengths from IRC on the AKARI satellite.
The geometric albedo and diameter data for approximately one-third of the asteroids in this new catalog are presented for the first time.
The AcuA-ISS catalog provides additional data that supplements previous asteroidal catalogs, such as SIMPS, AcuA, and {WISE}.

\bigskip
This work is based on observations with {AKARI}, a JAXA project with the participation of the ESA. 
We would like to express gratitude to staff of {AKARI} for their support in generating these observations. 
This study was partially supported by the Space Plasma Laboratory, ISAS, and JAXA as a collaborative research program.

\end{document}